# A radio detector array for cosmic neutrinos on the Ross Ice Shelf

Spencer R. Klein for the ARIANNA collaboration

*Abstract*—ARIANNA (The Antarctic Ross Ice Shelf Antenna Neutrino Array) is a proposed large detector for ultra-high energy (above $10^{17}$ eV) astrophysical neutrinos. It will instrument 900 km$^2$ of the 550 m thick Ross Ice Shelf (450 km$^3$ total volume) with radio detectors, to study the origins of ultra-high energy cosmic rays by searching for the neutrinos produced when these cosmic rays interact with the cosmic microwave background. Over 900 independently operating stations will detect the coherent radio Cherenkov emission produced when astrophysical neutrinos with energy above $10^{17}$ eV interact in the Antarctic Ross Ice Shelf. Each station will use 8 log periodic dipole antennas to look for short RF pulses, with the most important frequencies between 80 MHz and 1 GHz. By measuring the pulse polarization and frequency spectrum, the neutrino arrival direction can be determined. In one year of operation, the full array should observe a clear GZK neutrino signal, with different models predicting between 3 and 51 events, depending on the nuclear composition of the cosmic-rays and on the cosmic evolution of their sources.

## I. Introduction

THE origin of ultra-high energy (UHE) cosmic rays (CRs) is one of the great mysteries of modern physics. Despite 100 years of effort, we still do not know where these particles are produced. Both ground-based detector arrays and air fluorescence detectors have observed CR air showers with energies above $10^{20}$ eV, with the highest reported energy being $3\times10^{20}$ eV. These particles are a mixture of protons and heavier nuclei. Tight limits have been set on the photon fraction.

Because electrically charged nuclei bend in interstellar magnetic fields, even if we measure their arrival direction on the Earth, we cannot project them back to their sources. At lower energies, below $10^{16}$ eV, CRs likely have a galactic origin in supernova remnants [1]. However, at higher energies, there are no attractive galactic sources, and an extragalactic origin is preferred.

As Greisen, Zatsepin and Kuzmin (GZK) pointed out, protons with energies above $4\times10^{19}$ eV are likely to interact in flight, so the sources cannot be too distant [2, 3]. These protons are sufficiently Lorentz boosted that they can interact with the $3^0$K cosmic microwave background radiation (CMBR), forming a $\Delta^+$ resonance. The $\Delta^+$ decays to $p\pi^0$ or $n\pi^+$. The subsequent $\pi \rightarrow \mu\nu_\mu$ and $\mu \rightarrow e\nu_e\nu_\mu$ decays lead to three neutrinos (here, we do not distinguish between neutrinos and antineutrinos) plus an electron; the proton survives, but with a lower energy (neutrons will decay back into protons). This process limits the range of protons with energies above $4\times10^{19}$ eV to about 50 Mpc. If UHE CRs are heavier nuclei, then they will be photodissociated by the CMBR process, leading to a similar range limit. UHE photons face even more stringent limitations because they quickly pair-convert on the CMBR.

At low energies, the cosmic-ray arrival directions are expected to be largely isotropic, because the galactic magnetic fields randomize the directions of charged cosmic rays. At ultra-high energies, the GZK range limit could introduce some anisotropy, but this has not been observed [4]. However, at lower energies, where the flux is larger and the experimental sensitivity is higher, a small anisotropy has been observed [5]; this is not yet understood.

The same process that limits the range of UHE cosmic rays also offers an opportunity to study their existence: the neutrinos produced by GZK interactions travel in straight lines, and can easily traverse cosmic distances. Neutrinos from pion (or kaon) decay are produced in a 2:1 $\nu_\mu:\nu_e$ ratio. Over cosmic distances, they oscillate and arrive at Earth in a $\nu_\mu:\nu_e:\nu_\tau$ ratio of 1:1:1. GZK neutrinos typically have energies between $10^{17}$ and $10^{20}$ eV. There is also a somewhat smaller flux of neutrinos from neutron beta decay, with energies between $10^{16}$ and $10^{17}$ eV. Because of the lower energy and lower event rates, most efforts have focused on the higher energy peak.

Since the UHE cosmic ray flux, CMBR flux and the interaction cross-sections are well known, the GZK neutrino flux may be reliably calculated, subject to assumptions about

Manuscript received June 15, 2012. This work was funded in part by the U.S. National Science Foundation under grant number 0969661 and the Department of Energy under contract number DE-AC-76-00098.

Spencer R. Klein is with Lawrence Berkeley National Laboratory and the University of California, Berkeley, CA, 94720 (phone (510) 486-5470, email: srklein@lbl.gov).

the cosmic ray composition and the cosmic evolution of UHE sources. The latter consideration determines how source density changes at redshifts up to ~ 3-4 [3]. Calculations indicate that a detector with a 100 km$^3$ active volume is needed to observe a signal of 100 neutrinos in 3-5 years [6]. The largest optical Cherenkov detector, IceCube, has an active volume of about 1 km$^3$ [7], and the partially completed detector has not yet observed any GZK neutrino candidates [8]. The proposed KM3Net detector may reach 5 km$^3$ [9]. However, because of the limited absorption and scattering lengths in both ice and water, a 100 km$^3$ optical Cherenkov detector would require a prohibitively high number of sensors. For this reason, a new technology is needed. Although acoustic detection of showers has been considered, it suffers from short attenuation lengths in ice. The Auger and Fly's Eye collaborations have looked for air fluorescence signals of neutrino interactions in the atmosphere or the rock below. An enormous detector (far larger than the Auger array) would be required to see a reasonable GZK signal. Radio-detection is an attractive choice.

## II. NEUTRINO INDUCED SHOWERS AND RADIODETECTION OF NEUTRINO SHOWERS

Different types of neutrino interactions produce rather different topologies. Neutral current (NC) interactions transfer energy to a target nucleon, producing a hadronic shower. $\nu_\mu$ charged current (CC) interactions create a hadronic shower plus a high-energy muon. $\nu_e$ CC interactions produce an electromagnetic shower from the produced electron, plus a hadronic shower from the target nucleon. The electromagnetic and hadronic showers develop quite differently, because, at energies above $10^{16}$ eV, the electromagnetic shower is greatly elongated by the Landau-Pomeranchuk-Migdal (LPM) effect [10]. At energies above $10^{20}$ eV, photonuclear and electronuclear interactions overshadow bremsstrahlung and pair production, and the differences stop growing [11]. $\nu_\tau$ CC interactions produce a variety of topologies, depending on how the $\tau$ decays. At GZK energies, the differences between electromagnetic and hadronic showers are small to moderate.

Radiodetection is most sensitive to $\nu_e$ CC interactions, since they deposit all of their energy in a compact shower. It can also detect the hadronic showers produced by all-flavor NC interactions and from the hadronic showers in $\nu_\mu$ and $\nu_\tau$ CC interactions.

Coherent Cherenkov emission occurs when the wavelength of the emitted photon is larger than the transverse size of the shower, so that the Cherenkov amplitudes from each shower particle add in-phase. The radiation also includes amplitudes due to starting and stopping radiation from the produced charged particles; the latter is essentially bremsstrahlung [12]. In essence, at long wavelengths, the shower acts as a single particle with a net charge given by the charge excess in the shower [13]. The coherent amplitude is large because, near the tail end of the shower, there is a significant (~20%) excess of electrons over positrons. This excess comes about because photons can Compton scatter atomic (target) electrons into the beam and because shower positrons can annihilate on target electrons.

The characteristics of this radiation have been scrutinized carefully, both theoretically and experimentally. Radiation from this excess is centered at the Cherenkov angle, about 56 degrees for radio frequencies in ice. At the Cherenkov angle, the electric field amplitude increases linearly with frequency, up to a cutoff frequency $f_0$ ~ 1 GHz, where the wavelength becomes smaller than the transverse shower size. The width of the angular distribution depends on the frequency. At the cutoff frequency, the radiation forms a very thin cone around the Cherenkov angle. At lower frequencies, the angular distribution widens, and, at low frequencies $f << f_0$, the radiation is nearly isotropic. The relationship between frequency and angular width comes from the shower length. When the LPM effect reduces cross-sections enough to lengthen the shower, it also narrows the angular spread. The angular width of the peak, $\Delta\theta$, at a frequency $f$ is roughly [14]:

$$\Delta\theta \sim 2.7^o \left(\frac{f_0}{f}\right) \left[\frac{E_{LPM}}{0.14E + E_{LPM}}\right]^{0.3}.$$

The LPM effect also introduces significant event-to-event variations in shower development, leading to corresponding variations in radio emission, but this equation gives a good average width. This frequency dependence is useful in determining the direction of detected neutrinos.

The radiation is linearly polarized in the plane that contains the shower direction and the direction of the Cherenkov photon.

Radio Cherenkov emission has been studied extensively in a series of experiments at the Stanford Linear Accelerator Center, where a beam of ~$10^{10}$ 25-GeV electrons was directed into blocks of ice, sand, and salt [15].

## III. RADIODETECTION EXPERIMENTS

Radio-detectors comprise two or three elements: the active target volume, the antennas and data acquisition system, and, sometimes, an intervening medium between the active volume and the antennas. For example, in experiments which use the Moon as a target, the active volume is 380,000 km away from the antenna. Experiments may be roughly classified by this separation distance; the larger the separation, the higher the threshold. They may also be classified by their choice of target medium.

Many radio telescopes have searched for radiation from neutrino interactions in the Moon [6]. Current experiments have energy thresholds far above $10^{20}$ eV, so are not relevant for GZK neutrino searches. Even the proposed South African/Australian square kilometer radio telescope array will have a threshold somewhat above $10^{20}$ eV. The FORTE (Fast On-orbit Rapid Recording of Transient Events) satellite searched for radio pulses emitted from the Greenland ice pack; it also had a very high energy threshold – above $10^{22}$ eV [16]. For these experiments, the intervening medium includes the ionosphere, which introduces significant signal dispersion.

Moving down in both separation distance and energy, the (Antarctic Impulsive Transient Antenna ANITA) balloon experiment twice circled Antarctica at an altitude of 35 km [17]. Its 40 (32 in the first flight) horn antennas searched for radiation in the Antarctic ice cap, up to 650 km from the detector, leading to a threshold around $10^{20}$ eV. From the two flights, they ended up with 1 candidate event, compatible with the 1± 0.4 event estimated background. From this, they set the limit shown in Fig. 1.

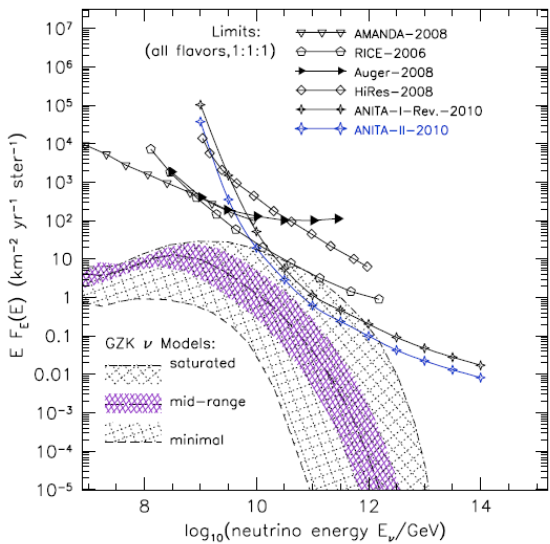

Fig. 1. Neutrino flux limits from the two ANITA balloon flights and other experimental results, compared with a range of theoretical predictions. From Ref. [17].

To reach a threshold of $10^{17}$ eV, it is necessary to locate the antennas in the active volume. For this, the preferred medium is Antarctic ice, which is available in large volumes, and has good radio-frequency electrical properties. This approach was pioneered by the Radio Ice Cherenkov Experiment (RICE) [18] which used Antarctic ice at the South Pole for its active medium. RICE placed 17 radio receivers at depths up to 350 m in holes drilled for the Antarctic Muon And Neutrino Detector Array (AMANDA) optical detectors. In addition to producing the first experimental result, the RICE collaboration was also the first to address some of the experimental problems of an ice-based detector.

One issue is the lack of a good electrical ground, since the nearest conducting Earth is kilometers away. This introduces many practical complications.

Another issue concerns signal propagation in the ice. The top ~ 100 m of the Antarctic icepack experiences a slow transition from solid ice at the bottom to packed snow at the top, known as the firn. The depth-dependent density leads to a depth-dependent index of refraction, causing radio waves to bend downward in the firn. It has also been proposed that surface waves may propagate in the ice-air interface [19].

Two experiments are currently proposing to deploy large detector arrays in the Antarctic Ice. The Askaryan Radio Arra (ARA) collaboration is proposing to build a large detector near the IceCube array at the South Pole [20]. ARIANNA is proposing a large array in Moore's Bay, on the Ross Ice Shelf.

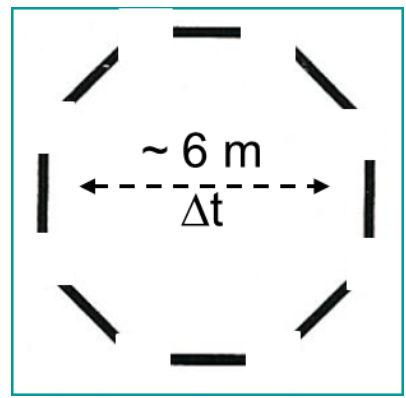

Fig. 2. Diagram of the ARIANNA antennas, looking downward. Δt schematically indicates the time difference between two parallel antennas, important for determining the direction to the neutrino interaction.

## IV. ARIANNA

The Antarctic Ross Ice-shelf ANtenna Neutrino Array (ARIANNA) takes advantage of some of the unique features of Moore's Bay [21]. There, 575 m of ice floats atop the Ross Sea. Moore's Bay is not near any glacial inflows, and radar and sonar surveys have shown that the ice-water interface is very flat [22]. So, the interface acts as a high-quality mirror (the loss is less than 3 dB) for reflecting radio waves. A downward-going neutrino that interacts in the ice will produce downward-going Cherenkov photons which are then reflected up to the surface detectors. This gives ARIANNA a unique sensitivity to downward-going neutrinos.

The full array will consist of an array of 900 stations, arranged on a 1 km grid. Each station will operate largely independently, collecting data autonomously using a single-station trigger.

ARIANNA detects radio waves using 8 log-periodic dipole antennas (LPDAs) - The antennas are buried pointing downward in an octagonal arrangement, as is shown in Fig. 2. The distance across the octagon is about 6 m. This arrangement allows for the measurement of neutrino direction for events observed by a single station, as will be discussed later.

The ice properties at the site were studied in 2007 by bouncing radio signals off the ice-water interface, using quad-ridged horn and Yagi antennas [23]. With this approach, there is an ambiguity between absorption at the ice-water interface, and absorption in the ice. Figure 3 shows the average attenuation lengths using two different surface antennas, as a function of frequency. Averaged over all ice depths, the attenuation lengths are 300-500 m, decreasing slowly with increasing frequency.

The absorption length of radio waves in the ice increases as the temperature drops. This accounts for most of the variation with depth. The temperature varies from about -25$^0$C just below the surface, to just below freezing at the bottom of the ice. Although the surface temperatures vary seasonally, a few meters deep, the ice temperature is constant year round. The measured absorption length is shorter than that observed at the South Pole [20], where the ice is colder.

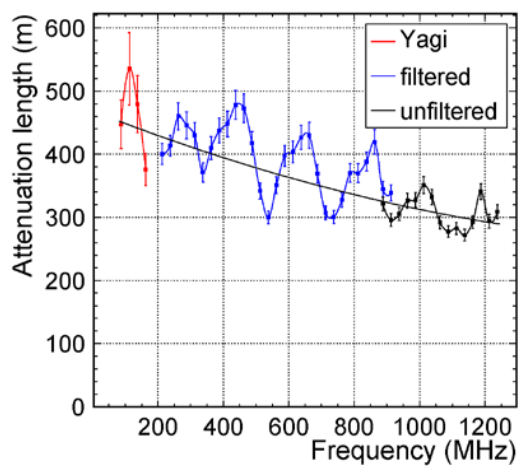

Fig. 3. Measured absorption length vs. frequency for radio signals bouncing off the ice-water interface. From Ref. [23].

The 183 MHz periodicity that is visible in Fig. 3 is not fully understood, but it could be related to interference between multiple signal paths (i.e. reflections). The details of the periodicity varied with the setup, but never disappeared. This points to a surface effect as the cause. More recently, studies have been done using a surface transmitter separated by 1 km from the receiver; this confirmed the rough attenuation properties shown above [24].

## V. ARIANNA HARDWARE

The ARIANNA LPDAs are essentially old-style television antennas. They have a fairly flat frequency response from 105 to 1300 MHz in air. As will be shown below, their response is modified in ice. They are specified to have 7-8 dBi forward gain in free air, with half power angles of 60-70$^0$ in the E-plane and 110-130$^0$ in the H plane. They have 50 Ω impedance and a voltage standing wave ratio (VSWR) better than 2:1 in the frequency range of interest [25].

The antennas are buried in shallow holes, with their top element buried 15-25 cm below the surface, as is shown in Fig. 4. As snow accumulates, they will gradually be buried more deeply, and, over time, pressure will increase the snow density around the antennas. The snow accumulation has been about 75 cm/year.

The ice shelf density increases from about 0.36 g/cm$^2$ at the surface to 0.92 g/cm$^2$ in the deep ice (below 100 m) [26]. The real part of the dielectric constant ε' depends linearly on the density and also varies slowly with the temperature. We use ε'=3.18 for the pure ice at the 0 to -25$^0$C ice temperatures found in the Ross Ice Shelf. Then, ε'= 1.8 on the surface. It is likely that, after correcting for the different temperatures, the absorption length will also scale with the density.

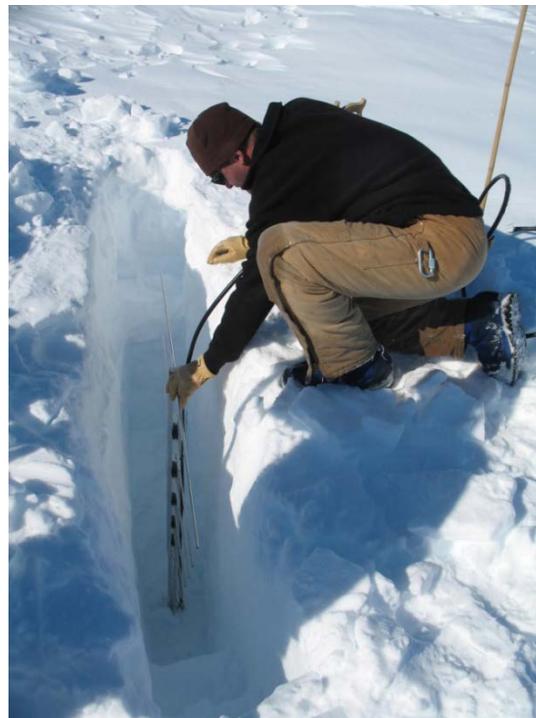

Fig. 4. One of the ARIANNA LPDA antennas being emplaced with the prototype station.

The electromagnetic properties of the ice have a couple of significant consequences. In the firn, the changing index of refraction causes the radio waves to refract, bending downward, away from the surface. Because of this, the ARIANNA stations are insensitive to neutrino interactions that occur near the surface, but which are distant from the station. Elsewhere, the refraction will need to be considered when determining the direction to the neutrino interaction, and the electric field polarization vector might be rotated.

The dielectric constant also alters the antenna performance, since a given frequency $f$ will correspond to a different wavelength: λ= c/f√ε'. It can also affect the antenna impedance. Figure 5 shows the measured voltage standing wavelength ratio (VSWR) for the ARIANNA antennas in free air and buried in the ice, plus two intermediate cases. The VSWR for the antennas is less than 2 for frequencies down to about 100 MHz in free air, but the low VSWR region extends down to 80 MHz when the antenna is buried in ice [27].

The antennas feed their signals to low-noise amplifiers based on (at least in the prototypes) Avago MGA-68243 GaAs MMIC (Monolithic Microwave Integrated Circuit) amplifiers. These amplifiers then feed a trigger and switched-capacitor array analog-to-digital converter [27, 28]. The station trigger will likely require 2 (or 3) of the antennas to be above a programmable threshold.

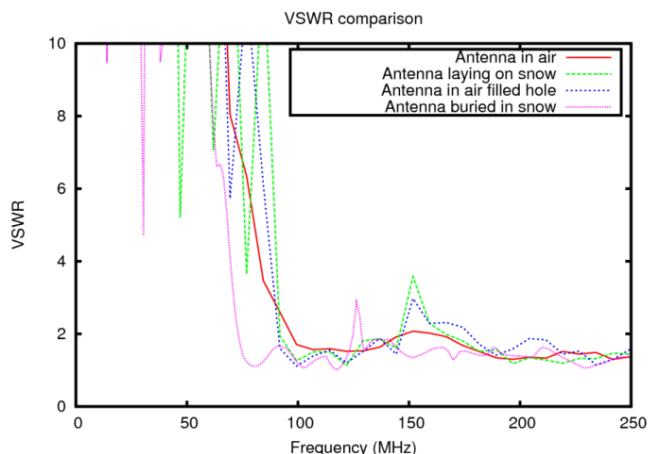

Fig. 5. The voltage standing wave ratio (VSWR) for the ARIANNA antennas in free air, with the antennas laying on the snow, in an air filled hole, and in a snow filled hole. From Ref. [27].

## VI. ANGULAR RESOLUTION AND EFFECTIVE VOLUME

It might seem surprising, but a single ARIANNA station is able to determine the arrival direction of a neutrino to within a few degrees [29, 30]. For this discussion, the neutrino-induced shower is treated as point-like. This assumption is somewhat questionable for nearby showers or very-high energy neutrinos, where the LPM effect lengthens the shower. However, with a more sophisticated treatment, the shower size should not lead to a significant degradation in resolution.

First, the direction from the station to the neutrino is determined. Using the time differences for the signal arrival, $\Delta t$, from opposite parallel antennas in Fig. 2, the direction from the station to the neutrino can be determined. A 0.5 nsec time difference with a 6 m separation leads to 0.5ns/30 ns ~ 16 mrad angular accuracy. There are two angles to be determined, but four antenna pairs, so the final resolution should be of order 10 mrad.

The neutrino direction is determined relative to the station-neutrino vector. Different methods are used to find the two relevant angles.

The first angle is determined from the frequency spectrum of the observed Cherenkov radiation. The spectrum depends on the angular distance between the observer and the Cherenkov cone – for an observer exactly on the cone, the radiation extends to the maximum cutoff frequency, while for an observer off the cone, the cutoff is at lower frequencies, as is discussed in Section II.

Although there is a slight spectral difference depending on if one is 'inside' the Cherenkov cone or 'outside' it, in a practical analysis, this difference is not observable, so there is a two-fold ambiguity regarding direction, except for signals observed exactly at the Cherenkov angle. The angular distance to the Cherenkov cone is often larger for closer (i.e. directly observed) events; as this angular distance increases, the resolution degrades.

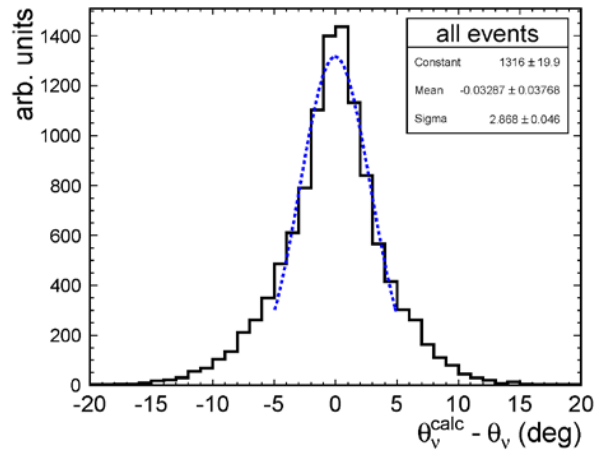

Fig. 6. ARIANNA zenith angle resolution for GZK neutrinos following the ν energy spectrum of Ref. [3]. This histogram is the result of a simulation, and the Gaussian curve is a fit to it. The mean resolution from the Gaussian is 2.9 degrees. The azimuthal angular resolution is similar. From Ref. [30].

The second angle comes from measuring the polarization of the radiation. The relative signal from the eight antennas can determine the polarization direction, relative to the vertical.
There is also a two-fold ambiguity here, since we cannot tell if the electric field is pointing up or down. If only four antennas were used, there would be an additional two-fold ambiguity present.

So, with eight antennas, there is a four-fold ambiguity, while four antennas lead to an eight-fold ambiguity. Many of the ambiguous cases can be eliminated, since the Earth absorbs $10^{17}$ eV neutrinos, so we do not expect any nearly vertical upgoing events.

There is an additional possible ambiguity, in that a single station cannot measure the distance to the event, and so cannot separate direct radiation from that which has reflected from the interface. Of course, in the best cases, we will observe both direct and reflected radiation, giving some information about the distances, and, therefore the neutrino energy.

The angular resolution can be estimated with a simple toy model. It uses a simple parameterization of the cascade generation in ice, as a function of frequency [31]. The signals are then propagated to the detector, ignoring the effects of the firn. The antenna and electronics bandwidth is taken to be 100 MHz – 1 GHz, and electronic noise is neglected. The signal is assumed to be a point source and near-field effects are neglected. This model gives a mean overall angular resolution of 1.7 degrees.

A more detailed simulation, which includes noise, a detector trigger, and signal propagation in the firn finds a somewhat worse angular resolution, as is shown in Fig. 6. This is for neutrinos with a GZK spectrum, arriving at Earth isotropically. The zenith angle resolution is particularly important, since a measurement of the neutrino flux as a function of zenith angle can be used to determine the neutrino-nucleon cross-section, by measuring the neutrino absorption in the Earth [32].

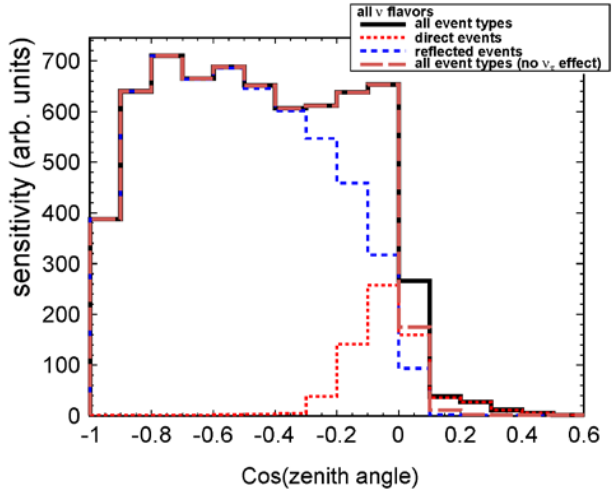

Fig. 7. The ARIANNA angular acceptance for direct signals (dotted red line) and for signals reflected off the ice-water interface (dashed blue line), as a function of the cosine of the zenith angle. The black line shows the sum of all events; it is somewhat hidden by the dashed blue line. The dashed brown line is the same, but without the effect of $\nu_\tau$ regeneration. Most of the direct signal is near the horizon, while the reflected events are seen over all downgoing zenith angles. From Ref. [30].

For the events detected by a single station (the vast majority of events), this analysis cannot determine the distance between the station and the neutrino interaction. Because of this, we cannot measure the neutrino energy on an event-by-event basis. We should be able to use the known detector response and statistical techniques to learn something about the overall energy spectrum.

## VII. PERFORMANCE

ARIANNA performance has been evaluated by using a series of Monte Carlo programs. The program generated neutrino events in the energy range $10^{17}$-$10^{21.5}$ eV in a 1:1:1 flavor ratio. Neutrino interactions in the Earth were included, as was $\nu_\tau$ regeneration. Hadronic and electromagnetic showers were produced from the charged and neutral current neutrino interactions. Tau double-bang events were simulated as two hadronic showers. The radio emission was parameterized. The LPM effect was included by narrowing the cone width, following Eq. (1). The resulting radio emission was propagated through the ice, taking into account the varying ice temperature and the firn density variation.

Radio absorption at the ice-water interface was taken to be -3 dB; the reflection was assumed to preserve polarization. The antenna was modeled using its known parameters, while the waveform digitizer and trigger performance were based on current prototypes.

Figure 7 shows the sensitivity as a function of zenith angle. Most of the events are observed via their reflections from the ice-water interface. These events extend from vertically downgoing to events coming from slightly below the horizon, with a fairly flat zenith angle distribution. Directly observed events are also seen, but only from a fairly narrow band near the horizon. The ratio of reflected to direct events reflects the performance increase from the reflective ice-water interface. Very few upward-going events are seen, because of neutrino absorption in the Earth.

Figure 8 shows the effective volume (at the trigger level) as a function of neutrino energy and flavor. The effective volume is the volume of an equally sensitive detector that is 100% efficient. Below $10^{18}$ eV, the effective volume is highest for $\nu_e$; at higher energies, it is highest for $\nu_\tau$. The higher $\nu_\tau$ effective volume is due to $\nu_\tau$ regeneration: $\nu_\tau$ travelling through the Earth may interact, producing a τ which then decays into a lower energy $\nu_\tau$ with lower energy. A high energy $\nu_\tau$ can traverse a thicker absorber than the other flavors, and emerge as a lower energy neutrino. So, $\nu_\tau$ are detectable at zenith angles at which $\nu_e$ would be absorbed. The flavor-averaged effective volume is about 0.5 km$^3$/station at $5\times10^{18}$ eV, the peak of the GZK ν spectrum curve.

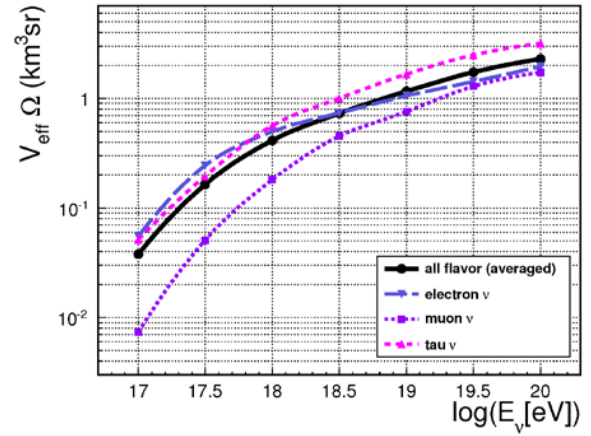

Fig. 8. The effective volume of a single ARIANNA station, as a function of energy, for the three neutrino flavors. This is averaged over incident angles. From Ref. [30].

Based on this sensitivity, different theoretical models predict that the full array should observe between 12 and 51 events per year [30] – assuming that the highest energy cosmic-rays are mostly protons. The Engel, Seckel and Stanev calculation [3] predicts 35 events/year. Another calculation takes the primaries to be mostly heavier nuclei [33], leading to a lower neutrino flux, and a prediction of 3 events/year. These rates conservatively assume that the array operates only during summer months, when solar power is available.

## VIII. POWER

The current prototypes are powered by solar cells and wind generators. They draw about 30 W in full operation; this will be reduced to considerably less than 10 W in future versions. The solar cells provide ample summer power, but are useless during the Antarctic winter. The wind generators have, so-far, not been able to provide adequate winter power, indicating that there is little winter wind at the site (matching the summer observations). We are evaluating possible future power systems, including enhanced wind, ultra-low power operations, and/or only operating for part of the year.

## IX. CURRENT STATUS

Fieldwork has been ongoing at the ARIANNA site since 2007, with yearly campaigns since 2009. An initial 4-antenna prototype station was deployed in 2009 [27], and a second prototype (with more advanced electronics) was deployed in late 2011. A 6-7 station prototype array will be deployed over the next two years (2012-14).

In addition to the studies of ice properties, we have learned much about the site and the necessary environmental characteristics of the hardware. We have estimates of the annual temperature profiles, wind speeds, and snow accumulation rates. The winds have been lower than predicted. We have also searched for anthropogenic noise at the site, and, except for a now-remediated switching power supply in the internet repeater, have not seen any. The detector triggers appear consistent with thermal noise. Minna Bluff serves as an excellent RF shield, blocking radio waves from McMurdo station.

Because of the importance of being able to adequately monitor and control the system in the absence of local (human) intelligence, year-round internet access becomes important. As a step toward that goal, we have established a seasonal internet repeater on Mt. Discovery, where it has line-of-sight access to both the ARIANNA site and to McMurdo station.

## X. CONCLUSIONS

The observation of ultra-high energy cosmic neutrinos will help answer a compelling physics question: where are the ultra-high energy cosmic ray accelerators in the universe? This question can be answered with a large (450 km$^3$) array of radio-detection stations. Radio-detection is quickly becoming a mature technique, and prototype Antarctic in-ice detectors are already in operation. The ARIANNA site on the Ross Ice Shelf provides a 570 m thick slab of ice with good transmission qualities for radio waves. The ice-water interface provides low-loss reflection of radio waves, enabling the detection of downward-going neutrinos. A 6-7 station "Hexagonal Radio Array" will be deployed in the next few years; this array should be well-poised to observe GZK neutrinos.

ACKNOWLEDGMENT

We thank Stijn Buitink for the 'toy model' analysis of the angular resolution. We thank Raytheon Polar Services Corporation and the entire crew at McMurdo Station for their excellent logistical support.